# Large magnetoresistance dips and perfect spin-valley filter induced by topological phase transitions in silicene

Worasak Prarokijjak[a], Bumned Soodchomshom[a,*]

[a]Department of Physics, Faculty of Science, Kasetsart University, Bangkok 10900, Thailand

*Corresponding author

Email address: Bumned@hotmail.com; fscibns@ku.ac.th

**Abstract**

Spin-valley transport and magnetoresistance are investigated in silicene-based N/TB/N/TB/N junction where N and TB are normal silicene and topological barriers. The topological phase transitions in TB's are controlled by electric, exchange fields and circularly polarized light. As a result, we find that by applying electric and exchange fields, four groups of spin-valley currents are perfectly filtered, directly induced by topological phase transitions. Control of currents, carried by single, double and triple channels of spin-valley electrons in silicene junction, may be achievable by adjusting magnitudes of electric, exchange fields and circularly polarized light. We may identify that the key factor behind the spin-valley current filtered at the transition points may be due to zero and non-zero Chern numbers. Electrons that are allowed to transport at the transition points must obey zero-Chern number which is equivalent to zero mass and zero-Berry's curvature, while electrons with non-zero Chern number are perfectly suppressed. Very large magnetoresistance dips are found directly induced by topological phase transition points. Our study also discusses the effect of spin-valley dependent Hall conductivity at the transition points on ballistic transport and reveals the potential of silicene as a topological material for spin-valleytronics.

**Keywords:** Silicene; Topological phase transition; spin-valley filter; Magnetoresistance;

## 1. Introduction

After the discovery of graphene [1], two-dimensional (2D) materials beyond graphene have drawn great interest in the field of condensed mater [2, 3]. Silicene, a novel two-dimensional silicon allotrope akin to graphene, has been both theoretically predicted [4-7] and experimentally synthesized [8-13]. It has become one of promising materials for modern electronic devices, such as spin-valleytronics [14-16]. The first silicene-based field effect transistor operated at room temperature has been recently fabricated [11]. Silicene is a monolayer of silicon with atoms arranged in honeycomb lattice. Its atomic structure is buckled related to mixed $sp^2 - sp^3$ hybridizations [17] and has strong spin orbit interaction (SOI) [6], unlike that in graphene which is planar and has weak spin orbit interaction [18]. Buckled structure leads to tunable energy gap by perpendicular electric field [19-20], due to A- and B-sublattices placed in different positions. The carriers in silicene are governed by Dirac fermions with spin-valley-dependent mass controlled by external fields. The presence of SOI may give rise to quantum spin Hall (QSH) Effect which was firstly proposed by Kane and Mele [21] in graphene including effect of intrinsic SOI. Unfortunately, the subsequent work found that in SOI in graphene is rather weak [18]. In contrast to graphene, strong SOI in silicene leads to prediction of rich phase [6, 15, 22-26]. It undergoes a topological phase transition from QSH state, 2D topological insulator, to a trivial (or band) insulator, quantum valley Hall (QVH) insulator, by varying perpendicular electric field [22]. Quantum anomalous Hall (QAH) Effect occurs in silicene induced by magnetization and SOI [23, 24]. Spin polarized quantum anomalous Hall (SQAH) insulator and anti-ferromagnetic (AF) phase are induced by interplay of electric field and magnetization [15]. The various types of topological phase transitions are classified by spin-valley Chern numbers due to spin and valley degrees of freedom in silicene. The trivial insulator phases are related to the first and spin-Chern numbers $(C, C_s)$ are zeroes $(0, 0)$ while topological insulator phases occur when $(C, C_s)$ are not zeroes [25]. Quantum spin-valley Hall conductivities may be given associated with spin-valley Chern numbers.

Recently, ballistic spin-valley transport properties in silicene junctions have been investigated by several works [16, 27-34]. Charge transport in $pn$ and $npn$ junctions in silicene were investigated, to show the conductance being almost quantized 0, 1 and 2 [27]. Spin-valley polarized currents have been investigated [28-30]. Defect enhanced spin and valley polarizations are possible in silicene superlattices [28-29]. The spin-valley-polarized Andreev reflection at the interface silicene-based normal/superconducting junction is found to be fully controlled by external electric field [30]. Perfect spin-valley filtering controlled by electric field has been proposed in a ferromagnetic silicene junction [16] when A and B sublatices are induced into ferromagnetism by different exchange fields. Control of spin-valley currents by circularly polarized lights have also been investigated [31-32], since it can induce valley-dependent-Dirac mass into silicene [33].

In this paper, we investigate spin-valley currents and magnetoresistance in silicene-based N/TB/N/TB/N junction where N and TB are normal silicene and topological-phase-transition barriers, respectively. We assume that in the two TB-barriers, perpendicular electric field [22], staggered exchange field [15,16] and circularly polarized lights are applied [33]. The effect of topological phase transitions in silicene on spin-valley transport and magnetoresistance is the main objective of our work. The effect which is directly due to topological phase transitions in the barriers has not been studied by previous works. Spin-valley quantum Hall conductivities in barriers are taken into consideration. The topological phase transition in the barriers can be tuned by varying electric field, exchange energy and frequency of circularly polarized light, controlling species of electron carriers to transport in the junctions.

## 2. Hamiltonian model

Let us first consider the tight-binding Hamiltonian in our model, a silicene-based N/TB/N/TB/N junction, as seen in Fig.1a. In TB regions, it may be modeled as of the form [15]

$$h_{TB} = -t \sum_{<i,j>\alpha} c_{i\alpha}^{\dagger} c_{j\alpha} + i\frac{\Delta_{so}}{3\sqrt{3}} \sum_{<<i,j>>\alpha\beta} \nu_{ij} c_{i\alpha}^{\dagger} \sigma_{\alpha\beta}^{z} c_{j\beta} - i\frac{2}{3}\Delta_{R} \sum_{<<i,j>>\alpha\beta} \gamma_i c_{i\alpha}^{\dagger} \left(\vec{\sigma} \times \hat{\xi}_{ij}\right)_{\alpha\beta}^{z} c_{j\beta}$$

$$+ \sum_{i\alpha} \gamma_i e\ell E_z c_{i\alpha}^{\dagger} c_{i\alpha} + \sum_{i\alpha\beta} \gamma_i M c_{i\alpha}^{\dagger} \sigma_{\alpha\beta}^{z} c_{i\beta} + i\frac{\Delta_{\Omega}}{3\sqrt{3}} \sum_{<<i,j>>\alpha\beta} \nu_{ij} c_{i\alpha}^{\dagger} c_{j\beta} + \sum_{i\alpha} U c_{i\alpha}^{\dagger} c_{i\alpha},$$

$$(1)$$

where $c_{i\alpha}^{\dagger}$ ($c_{j\beta}$) is creation(destruction) field operator at site i(j) for electron with spin polarization $\alpha(\beta)$ and $<i,j>$ ($<<i,j>>$) run over all the nearest-neighbor (next-nearest neighbor) hoping sites. The first term represents graphene-like Hamiltonian with hoping energy $t = 1.6$ meV for silicene. The second term represents spin-orbit interaction effect in silicene with $\Delta_{so} = 3.9\,\text{meV}$ [34] where $\vec{\sigma} = <\sigma^x, \sigma^y, \sigma^z>$ is vector of Pauli spin matrix acting on real-spin states. $v_{ij} = 1(-1)$ if the next-nearest-neighboring hoping is anticlockwise (clockwise) respect to direction normal to silicene sheet. The third term represents Rashba spin orbit interaction with $\Delta_R = 0.7\,\text{meV}$ where $\gamma_i = 1(-1)$ for A-(B-) sublattice and $\hat{\xi}_{ij} = \vec{\xi}_{ij} / |\vec{\xi}_{ij}|$ with $\vec{\xi}_{ij}$ connecting two site i and j in the same sublattice. The fourth term represents interaction due to applying perpendicular electric field $\Delta_E = e\ell E_Z$ where $e, E_z$ and $\ell$ are bare electron charge, electric field and buckling parameter, respectively. The fifth term represents interaction induced by staggered exchange field with exchange energy $\Delta_M = M$ which may be realized by depositing magnetic insulators with exchange energy of M on top and bottom of silicene sheet with different exchange field directions [16, 23]. The sixth term represents interaction induced by off-resonant circularly polarized light irradiated onto silicene sheet with vector potential $\vec{A} = \Lambda < \pm\sin\Omega t, \cos\Omega t, 0 >$, where $\Lambda$ and $\Omega$ are the amplitude and frequency of light respectively. This is to get $\Delta_\Omega = \pm\hbar v_F^2 A^2 / a^2\Omega$ where $+(-)$ denotes right (left) circulation. $A = ea\Lambda / \hbar$, $a = 3.86\,\overset{0}{\text{A}}$ and $v_F = 5.5\times10^5\,\text{m}/\text{s}$ are dimensionless amplitude [33], the lattice constant and the Fermi velocity [34], respectively. The last term represents chemical potential applied by gate voltage with electric potential $-U / e$. In the low energy limit, the effect of Rashba term is usually neglected. Therefore, low energy effective Hamiltonian used to describe motion of quasiparticles, Dirac fermions, in TB-regions, related to tight-binding model in eq.1 may be given as

$$H_{TB} = \eta v_F \hat{p}_x \tau^x + v_F \hat{p}_y \tau^y + \left(\eta s\Delta_{so} + \eta\Delta_\Omega + s\Delta_M - \Delta_E\right)\tau^z + U, \qquad (2)$$

where $\vec{\tau} = <\tau^x, \tau^y, \tau^z>$ is vector of Pauli spin matrix acting on pseudo(or lattice)-spin states, $s = +(-)$ stands for electron with real-spin $\uparrow(\downarrow)$, $\eta = +(-)$ stands for electron in $k - (k'-)$ valley, and $\hat{p}_{x(y)} = -i\hbar\partial_{x(y)}$ is the momentum operator. In N-regions, there are no electric, exchange fields and circularly polarized light applied into silicene sheet. By taking $\Delta_\Omega = \Delta_M = \Delta_E = \mu = 0$ into eq.2, thus we get low energy effective Hamiltonian in NM-regions of the form

$$H_N = \eta v_F \hat{p}_x \tau^x + v_F \hat{p}_y \tau^y + \eta s\Delta_{so}\tau^z. \qquad (3)$$

It is seen that from eqs.2 and 3, in N-region the band dispersion is not spin-valley dependent

$$E_{\eta s} = \pm\sqrt{v_F^2 p_x^2 + v_F^2 p_y^2 + \Delta_{so}^2} \ , \qquad (4)$$

where $\pm$ denotes conduction(valence) band. In TB-regions, it is spin-valley dependent

$$E_{\eta s} = \pm\sqrt{v_F^2 p_x^2 + v_F^2 p_y^2 + \left(\eta s\Delta_{so} + \eta\Delta_\Omega + s\Delta_M - \Delta_E\right)^2} + U \ . \qquad (5)$$

The energy gap and Dirac mass in N-regions are $E_{gap,\eta s} = 2\Delta_{so}$ and $m_{\eta s} = \eta s\Delta_{so} / v_F^2$, respectively. Energy gap and Dirac mass in TB-regions are $E_{gap,\eta s} = 2\left|\eta s\Delta_{so} + \eta\Delta_\Omega + s\Delta_M - \Delta_E\right|$ and $m_{\eta s} = \left(\eta s\Delta_{so} + \eta\Delta_\Omega + s\Delta_M - \Delta_E\right) / v_F^2$, respectively. Plot of band structure in each region is shown in Fig1b.

## 3. Topological phase transitions in TB-regions

As mentioned in the first section, the topological phase transition is required in the TB-regions. The Fermi energy in TB-region must be inside the energy gap of the carriers, using the condition of $E = U$. The Fermi energy of electron only in N-regions lies above the gap leading to that all electron species can propagate through NM-regions and there is no topological phase in N-regions. Hence, TB-regions are considered as topological barriers in which phase transitions can be tunable by external forces. Let us start with spin-valley Chern numbers, $C_{\eta s}$, in the system that governed by Hamiltonian $H_{\eta s} = \vec{d} \cdot \vec{\tau}$. It is equivalent to the Pontryagin number which may be given by [25, 35, 36]

$$C_{\eta s} = \frac{1}{2\pi} \int d^2k \, \mathbb{F}_{\eta s} = \frac{1}{4\pi} \int d^2k \, \hat{d} \cdot \left( \frac{\partial \hat{d}}{\partial k_x} \times \frac{\partial \hat{d}}{\partial k_y} \right), \qquad (6)$$

where $\mathbb{F}_{\eta s}$ is spin-valley Berry's curvature. In TB-regions, we have $\vec{d} = <\eta v_F \hbar k_x, v_F \hbar k_y, \Delta_{\eta s}>$ and $\hat{d} = \vec{d} / \left|\vec{d}\right|$, with $\Delta_{\eta s} = \eta s\Delta_{so} + \eta\Delta_\Omega + s\Delta_M - \Delta_E$. Eq.6 gives rise to Chern numbers depending on the sign of Dirac mass of relativistic electron, as given by [25, 33, 35]

$$C_{\eta s} = \frac{\eta}{2} sgn(\Delta_{\eta s}) \ . \qquad (7)$$

Classification of topological phase in silicene may be given by using total Chern number $C$, real-spin Chern number $C_s$, valley Chern number $C_v$, and pseudo-spin Chern numbers $C_{ps}$, which are respectively defined as [25, 37]

$$C = C_{k\uparrow} + C_{k\downarrow} + C_{k'\uparrow} + C_{k'\downarrow},$$

$$C_s = \frac{1}{2}\left(C_{k\uparrow} + C_{k'\uparrow} - C_{k\downarrow} - C_{k'\downarrow}\right),$$

$$C = C_{k\uparrow} + C_{k\downarrow} - C_{k'\uparrow} - C_{k'\downarrow},$$

and

$$C_{ps} = \frac{1}{2}\left(C_{k\uparrow} + C_{k'\downarrow} - C_{k'\uparrow} - C_{k\downarrow}\right).$$

$$(8)$$

We note that the last term represents pseudo-spin Chern number. This is due to the fact that $\left|k\uparrow\right\rangle$ and $\left|k'\downarrow\right\rangle$ are wave functions of electron in A-sublattice equivalent to pseudo-spin up, while $\left|k\downarrow\right\rangle$ and $\left|k'\uparrow\right\rangle$ are wave functions of electron in B-sublattice equivalent to pseudo-spin down, clarified in refs. [38, 39]. For $(C, C_s) \neq (0,0)$, TB-regions may be considered as a topological barrier, where the cases of $(C, C_s) = (\pm 2, 0), (0, \pm 1)$ and $(\pm(or\mp)1, \pm(or\mp)1/2)$ are QAH, QSH and SQAH, respectively. For $(C, C_s) = (0,0)$, TB-regions may be considered as a trivial insulating barriers, where $(C_v, C_{ps}) = (\pm 1, 0)$ and $(0, \pm 1)$ yields QVH and quantum pseudo-spin Hall effect (QPSH), respectively (see Table I). When the Fermi level lies in the gap, the spin-valley Hall conductivity at zero temperature for this model may be obtained by the Kubo formulism [40-44], as given by

$$\sigma_{xy,\eta s} = \frac{e^2 \left(\hbar v_F\right)^2}{4\hbar \left(2\pi\right)^2} \int d^2k \frac{\eta \Delta_{\eta s}}{\left(\left(\hbar v_F k\right)^2 + \Delta_{\eta s}^2\right)^{3/2}}$$

$$= \frac{e^2}{2h}\left(\frac{1}{2\pi}\int d^2k \mathbb{F}_{\eta s}\right) = \frac{e^2}{2h}C_{\eta s}. \qquad (9)$$

This Hall conductivity derived by the Kubo formula can be represented in term of the Chern numbers induced by the Berry's curvature in momentum space TKNN formula [44]. The spin-valley Hall conductivity formula given by eq.9 is used to describe the spin-valley currents at the edge when topological phase occurs. The total, spin, valley and pseudo-spin Hall conductivities may be defined as

$$\sigma_{xy} = \sigma_{xy,k\uparrow} + \sigma_{xy,k\downarrow} + \sigma_{xy,k'\uparrow} + \sigma_{xy,k'\downarrow}, \qquad \sigma_{xy}^{s} = \sigma_{xy,k\uparrow} + \sigma_{xy,k'\uparrow} - \sigma_{xy,k\downarrow} - \sigma_{xy,k'\downarrow},$$

$$\sigma_{xy}^{v} = \sigma_{xy,k\uparrow} + \sigma_{xy,k\downarrow} - \sigma_{xy,k'\uparrow} - \sigma_{xy,k'\downarrow}, \text{ and } \sigma_{xy}^{ps} = \sigma_{xy,k\uparrow} + \sigma_{xy,k'\downarrow} - \sigma_{xy,k\downarrow} - \sigma_{xy,k'\uparrow},$$

$$(10)$$

respectively. The QSH yields $\left|\sigma_{xy}^{s}\right| = e^2 / h$ and $\left|\sigma_{xy}^{T}\right| = \left|\sigma_{xy}^{v}\right| = \left|\sigma_{xy}^{ps}\right| = 0$ [40]. The QVH yields $\left|\sigma_{xy}^{v}\right| = e^2 / h$ and $\left|\sigma_{xy}^{T}\right| = \left|\sigma_{xy}^{s}\right| = \left|\sigma_{xy}^{ps}\right| = 0$ [40]. The QPSH yields $\left|\sigma_{xy}^{ps}\right| = e^2 / h$ and $\left|\sigma_{xy}^{T}\right| = \left|\sigma_{xy}^{s}\right| = \left|\sigma_{xy}^{v}\right| = 0$.

## 4. Scattering process and transport formulae

The transport properties in a silicene-based N/TB/N/TB/N junction, depicted in Fig. 1a, are studied in this section. The Hamiltonians in eqs. 2 and 3 are adopted to describe the motion of the carriers in N- and TB-regions, respectively. The current is assumed to flow in the x-direction. Therefore, when electron is injected into the N/TB interface with angle $\theta$ and energy E, the wave function in each region may be obtained as

$$\psi_{N1} = [\begin{pmatrix} 1 \\ Ae^{i\eta\theta} \end{pmatrix} e^{ikx} + r_{\eta s} \begin{pmatrix} 1 \\ -Ae^{-i\eta\theta} \end{pmatrix} e^{-ikx}] e^{ik_{\parallel}y},$$

$$\psi_{TB1} = [a_{\eta s} \begin{pmatrix} 1 \\ B_{\pm}e^{i\eta\phi_{+}} \end{pmatrix} e^{iq_{+}x} + b_{\eta s} \begin{pmatrix} 1 \\ -B_{\pm}e^{-i\eta\phi_{+}} \end{pmatrix} e^{-iq_{+}x}] e^{ik_{\parallel}y},$$

$$\psi_{N2} = [c_{\eta s} \begin{pmatrix} 1 \\ Ae^{i\eta\theta} \end{pmatrix} e^{ikx} + d_{\eta s} \begin{pmatrix} 1 \\ -Ae^{-i\eta\theta} \end{pmatrix} e^{-ikx}] e^{ik_{\parallel}y},$$

$$\psi_{TB2} = [e_{\eta s} \begin{pmatrix} 1 \\ B_{\pm}e^{i\eta\phi_{\pm}} \end{pmatrix} e^{iq_{\pm}x} + f_{\eta s} \begin{pmatrix} 1 \\ -B_{\pm}e^{-i\eta\phi_{\pm}} \end{pmatrix} e^{-iq_{\pm}x}] e^{ik_{\parallel}y},$$

and

$$\psi_{N3} = t_{\eta s} \begin{pmatrix} 1 \\ Ae^{i\eta\theta} \end{pmatrix} e^{ikx+ik_{\parallel}y},$$

where

$$A = \frac{E - \eta s \Delta_{so}}{\eta \sqrt{E^2 - \Delta_{so}^2}}, \ B_{\pm} = \frac{(E - U) - (\eta s \Delta_{so} + \eta \Delta_{\Omega} \pm s \Delta_{M} - \Delta_{E})}{\eta \sqrt{(E - U)^2 - (\eta s \Delta_{so} + \eta \Delta_{\Omega} \pm s \Delta_{M} - \Delta_{E})^2}},$$

$$k = \sqrt{E^2 - \Delta_{so}^2} \cos\theta, \ k_{\parallel} = \sqrt{E^2 - \Delta_{so}^2} \sin\theta$$

$$q_{\pm} = \sqrt{(E - U)^2 - (\eta s \Delta_{so} + \eta \Delta_{\Omega} \pm s \Delta_{M} - \Delta_{E})^2} \cos\phi_{\pm},$$

with $\phi_{\pm} = \sin^{-1}\dfrac{\sqrt{E^2 - \Delta_{so}^2}\,\sin\theta}{\sqrt{\left(E - U\right)^2 - \left(\eta s\Delta_{so} + \eta\Delta_{\Omega} \pm s\Delta_M - \Delta_E\right)^2}}$ .

$$(11)$$

The notation $\pm$ in the wave function in the TB2-region represents P and AP-types of the junction for $+$ and $-$, respectively. The magnetizations in TB2 can be set to be parallel or anti-parallel to TB1. The coefficients, $r_{\eta s}$, $a_{\eta s}$, $b_{\eta s}$, $c_{\eta s}$, $d_{\eta s}$, $e_{\eta s}$, $f_{\eta s}$ and $t_{\eta s}$ can be determined by matching the wave function in eq.11 with the boundary conditions below

$$\psi_{N1}(0) = \psi_{TB1}(0) \ , \ \psi_{TB1}(d) = \psi_{N2}(d) \, ,$$

$$\psi_{N2}(d + L) = \psi_{TB2}(d + L) \ \text{ and } \ \psi_{TB2}(2d + L) = \psi_{N3}(2d + L) \, .$$

$$(12)$$

By doing this, the ballistic transmission in the model may be calculated by the usual formalism $T_{\eta s}(\theta) = \left|t_{\eta s}\right|^2$.

The conductances are related to the transmissions $T_{\eta s}(\theta)$. By using the Landauer formula [45] which integrates transmissions overall incident angle, the formula for the spin-valley dependent conductance $G_{\eta s}$ may be given by

$$G_{k\uparrow(\downarrow)} = \frac{1}{2}G_0 \int_{-\pi/2}^{\pi/2} d\theta \cos\theta\, T_{k\uparrow(\downarrow)}(\theta) \ \text{ and } \ G_{k'\uparrow(\downarrow)} = \frac{1}{2}G_0 \int_{-\pi/2}^{\pi/2} d\theta \cos\theta\, T_{k'\uparrow(\downarrow)}(\theta), \quad (13)$$

where $G_0 = \dfrac{4e^2}{h}N(E)$ is a conductance of non-impurity silicene with $N(E) = \dfrac{w}{\pi\hbar v_F}\sqrt{E^2 - \Delta_{so}^2}$ being the density of state for non-impurity silicene. The width of the junction is denoted as $w$. The net conductance of the junction is given by

$$G = G_{k\uparrow} + G_{k\downarrow} + G_{k'\uparrow} + G_{k'\downarrow} \, . \qquad (14)$$

The pseudospin- and actual (or real) spin- and valley- conductancese may be defined as

$$G^{ps} = G_{k\uparrow} + G_{k'\downarrow} - G_{k\downarrow} - G_{k'\uparrow}, \qquad G^s = G_{k\uparrow} + G_{k'\uparrow} - G_{k\downarrow} - G_{k'\downarrow}$$

and $\qquad G^v = G_{k\uparrow} + G_{k\downarrow} - G_{k'\uparrow} - G_{k'\downarrow} \, ,$

$$(15)$$

respectively. The magnetoresistance in this study is defined by

$$MR(\%) = \frac{G_{AP} - G_P}{G_{AP}} \times 100\% \qquad (16)$$

We note that this definition of MR is used to study the MR dips induced by topological phase transitions.

## 6. Results and discussion

In the numerical results, the topological barriers, TB-regions, may be achieved by taking $E = U$. This condition may give rise to the Fermi level in TB-regions inside energy gap to cause topological phase in TB-regions [40]. The topological phase in the TB-barriers may be classified by the set of spin-valley Chern numbers given in Table 1. We firstly show the spin-valley conductance in case of L=0. In this condition, the junction can be considered as like a N/TB/N junction with barrier thickness of 2d. When exchange filed and circular polarized light are not applied into the barrier, the TB-region becomes QSH for $-\Delta_{so} < \Delta_E < \Delta_{so}$ and the other regions are QVH (see Fig 2a). The transition points between QSH and QVH are at $\Delta_E = \pm\Delta_{so}$. These have been reported in several works [22, 33, 40]. It is found that, the junction exhibits strong insulator when QSH and QVH occur. This is because QSH and QVH are insulating in bulk leading to TB-region being strong insulator when topological phase occurs. Interestingly, at $\Delta_E = \pm\Delta_{so}$ which are the transition points, they are neither QSH nor QVH phases. The conductance of the junction are almost $G(\pm\Delta_{so}) = 2G_0$, considered as a very strong conductor. At $\Delta_E = \Delta_{so}$, the junction allows only electrons with $k\uparrow$ and $k'\downarrow$ to transport while at $\Delta_E = -\Delta_{so}$, it allows only electrons with $k\downarrow$ and $k'\uparrow$ to transport. When d increases (see Fig.2b), almost 100% pseudo-spin polarization occurs at the transition points (zero valley and spin polarizations at these points $G^s = G^v = 0$). Transition points between QSH and QVH lead to pseudo-spin filtering effect. We identify that the key factor behind the perfect pseudo-spin filter at the transition points may be related to spin-valley Hall conductivities or transverse conductivities given in eq.9 at the transition points, as obtained by

$$\sigma_{xy,\eta s} = \eta\frac{e^2}{4h}\text{sgn}\left(\eta s\Delta_{so} + \eta\Delta_\Omega + s\Delta_M - \Delta_E\right)$$

$$= \eta \frac{e^2}{4h} \text{sgn}(\eta s \Delta_{so} - \Delta_E).$$

The spin-valley ballistic conductances in our model may be related to longitudinal conductivity $G_{\eta s} \equiv \sigma_{xx,\eta s}$ where

$$\sigma = \begin{pmatrix} \sigma_{xx} & \sigma_{xy} \\ -\sigma_{xy} & \sigma_{yy} \end{pmatrix}$$

are conductivity tensor with $\sigma_{yy} = 0$ (current flowing only in x-direction). At the transition point $\Delta_E = \Delta_{so}$, we have $\sigma_{xy,k\uparrow} = \sigma_{xy,k\downarrow} = 0$ and $\sigma_{xy,k\downarrow} = -\sigma_{xy,k'\uparrow} = -\frac{e^2}{4h}$, while $G_{k\uparrow} = G_{k'\downarrow} \cong G_0$ and $G_{k\downarrow} = G_{k'\uparrow} = 0$. In contrast to the first case, at the transition point $\Delta_E = -\Delta_{so}$, it is given $\sigma_{xy,k\uparrow} = -\sigma_{xy,k'\downarrow} = -\frac{e^2}{4h}$ and $\sigma_{xy,k\downarrow} = \sigma_{xy,k'\uparrow} = 0$, while $G_{k\uparrow} = G_{k'\downarrow} = 0$ and $G_{k\downarrow} = G_{k'\uparrow} \cong G_0$. This is to say that emergence of Hall conductivity of electron at the TB-barrier may suppress its related longitudinal conductance, due to strong insulator in bulk but conducted at the edge. Electron may be allowed to perfectly transport when its related Hall conductivity vanishes at the TB-barrier. As understood, electron with spin "s" and valley $\eta$ would exhibit Klein tunneling [46] when it acquires zero mass $m_{\eta s} = \Delta_{\eta s} / v_F^2 = 0$. This is equivalent to $\sigma_{xy,\eta s} = 0$. Clearly, electrons with spin "s" and valley $\eta$ which are allowed to transport through the junction must obey the condition of $C_{\eta s} = 0$, while the other electrons may be completely suppressed when their spin-valley Chern numbers are not zero. This condition will be adopted to describe the perfect spin-valley filtering and very large magnetoresistance dips at the transition points in the next result.

We next study the spin-valley dependent conductances as a function of electric field when applying exchange field and circularly polarized light in the case of P-junction (see Figs.3a-3c). The junction is assumed to have $L = 0$. In Fig.3a, when the junction is applied only the staggered exchange field to get $\Delta_\Omega = 0$ but $\Delta_M \neq 0$, it is found that SQAH occurs as a buffer between QSH and QVH. The emergence of SQAH is directly due to applied exchange fields in the TB-region. In this case, it is also found that there are perfect spin-valley filter. Electron with spin "s" and valley "$\eta$" are allowed to transport at different magnitude of electric field, obeying the condition of $C_{\eta s} \equiv \mathrm{sgn}\left(\eta s \Delta_{so} + s \Delta_M - \Delta_E\right) = 0$. In Fig 3b, we study $G_{\eta s}$ when $\Delta_\Omega \neq 0$ but $\Delta_M = 0$. SQAH appears as a buffer state between QSH and QVH similar to the case of applied only exchange filed but different SQAH type. Four species of electrons are found to be perfectly split at specific electric field, predicted by the condition of $C_{\eta s} \equiv \mathrm{sgn}\left(\eta s \Delta_{so} + \eta \Delta_\Omega - \Delta_E\right) = 0$. In Fig 3c, we study $G_{\eta s}$ when $\Delta_\Omega = \Delta_M \neq 0$. It is found that peaks of $k_\downarrow$- and $k'_\uparrow$-electrons merge at $\Delta_E = -\Delta_{so}$ due to $C_{k\downarrow} = C_{k'\uparrow} = 0$, allowing two electron species transport at this point. Peaks of $k_\uparrow$- and $k'_\downarrow$-electrons are still distinguishable and are found at $\Delta_E = \Delta_{so} + 2\Delta_\Omega$ and $\Delta_E = \Delta_{so} - 2\Delta_\Omega$, respectively.

The conductance in the AP-junction and magnetoresistance are studied in Fig.4 when L=0. In Fig.4a, we consider $G_{\eta s}$ when $\Delta_\Omega = 0$ but $\Delta_M \neq 0$. Interestingly, the behavior of $G_{\eta s}$ resembles that in the P-junction when there is no applied exchange field (Fig. 4a is equal to Fig. 2a). The spin-valley filtering effect can be described using the case of non magnetic junction. This has been shown straightforwardly that Fig. 4c is equal to Fig.3b. The magnetoresistance are studied in Figs. 4b, 4d and 4f. It is found that large conductance dips occurs at the transition points of TB-barriers in the P-junctions. Magnitude of MR can be enhanced by increasing barrier thickness. Fig. 4e and Fig.4f show interesting result, when $\Delta_\Omega = \Delta_M = \Delta_{so}$. In this case, the currents can be controlled to flow by three species of electrons $k_\downarrow$-, $k'_\uparrow$ and $k'_\downarrow$-electrons at $\Delta_E = \Delta_{so}$ because of these electrons acquiring zero Chern numbers at this point, while only one electron specie $k_\downarrow$-electron are allowed to transport at $\Delta_E = 3\Delta_{so}$ because of only $C_{k\downarrow} = 0$. In the AP-junction, as we have discussed above its conductance spectra may be described by P-junction with $\Delta_M = 0$. Hence AP-junction in Fig.4e may be replaced by P-junction when $\Delta_\Omega = \Delta_{so}$ and $\Delta_M = 0$. Remember that The condition to get spin-valley filter is $C_{\eta s} \equiv \mathrm{sgn}\left(\eta s \Delta_{so} + \eta \Delta_{so} - \Delta_E\right) = 0$ thus we get peaks of $G_{k'\uparrow}$ and $G_{k\uparrow}$ at $\Delta_E = -2\Delta_{so}$ and $\Delta_E = +2\Delta_{so}$, respectively. Two peaks of $G_{k'\downarrow}$ and $G_{k\downarrow}$ are allowed to transport at the same point, $\Delta_E = 0$. We note that the filtering effect in AP-junction is not due to transition points, while spin-valley filtering effect in the P-junction is due directly to the transition points. This is because the spectra of conductance in The AP-junction are just described as equivalent to those in non-magnetic P-junction, not exactly the same junction. The topological phases in case of P-junction in Fig.4e, is described by the MR-dip behavior seen in Fig.4f. The very large MR dips are found between QVH and SQAH.

In Fig. 5, the spin-valley filtering effect and magnetoresistance for the case of $\Delta_\Omega = 0$ but $\Delta_M \neq 0$ have been investigated for various values of $L$. As seen in Fig.5a for P-junction, it is found that the spin-valley conductance peaks which are due only to transition points occurs when $L$ is small enough. When $L$ is large enough, the spin-valley conductance peaks which are not associated with phase transition appear, as seen in the case of $L = 100 \, \text{nm}$. Four peaks outside the region of $-10 \text{meV} \leq \Delta_E \leq 10 \text{meV}$ in Fig. 5a, these peaks may be generated related to the quantum interference inside N-barrier with thickness L. In the case of AP-junction, as we have discussed it can be considered as similar to that in the P-junction for no exchange fields. We can see the multiple peaks when increasing L to be large enough (see Fig.5b). The multiple conductance peaks arisen by increasing L gives rise to complicated MR dips. The numbers of MR dips are equal to the number of the conductance peaks of P-junction (see Fig.5c). This result of complicated conductance peaks found for large L may point out that very small L is required for the effect of spin-valley filter and MR dips generated by topological phase transitions. We note that, the giant magnetoresistance investigated in silicene system has been investigated by refs.[47] and [48]. The giant MR has been predicted in different structures. In ref.[47], MR in double magnetic strip-induced vector potentials to generate giant MR without considering influence of topological phase transition. In ref.[48], giant MR and perfect spin filter has been predicted in silicene-based nanoribbon.

Finally, the spin-valley and total conductances as a function of $\Delta_\Omega$ are investigated for the case of L=0 for P-junction, as seen in Fig.6. When $\Delta_E = \Delta_M = 0$, it is found that at the transition points $\Delta_\Omega = \pm \Delta_{so}$ between QSH and QAH, the junction exhibit $\mp 100\%$ real-spin polarization. Figs. 6b and 6c show perfect spin-valley filter due to topological phase transitions, when $\Delta_M$ is applied without electric field. QSH appears when $\Delta_M < \Delta_{so}$, while QPSH appears when $\Delta_M < \Delta_{so}$. The three species of electrons and single specie of electron can be controlled to transport by specific value of $\Delta_\Omega$ (as seen in Fig 6d). QSH and QPSH disappear, when $\Delta_E = \Delta_M = \Delta_{so}$.

## 7. Summary and conclusion

We have investigated spin-valley transport in silicene-based N/TB/N/TB/N junction where N and TB are normal silicene and topological barriers. In topological barriers, the Fermi energy lies inside the gap to get topological phase transition tunable by external forces, electric field, exchange field, and circularly polarized light. We showed that perfect spin-valley filter occurs at the topological transition points. Electrons that are allowed to transport at the transition points must obey the condition of zero-Chern number induced by zero-Berry's curvature which is equivalent to electron acquiring zero mass. In this regime, the massless Dirac electron may carry high ballistic conductance due to Klein tunneling without back reflection at normal angle of incidence. At the transition points, although the massive carriers with non-zero Chern number are fully suppressed in TB-regions, they may generate currents at the edge related to spin-valley quantum hall conductivity. We also showed that the four electron species, $I_{k\uparrow}, I_{k\downarrow}, I_{k'\uparrow},$ and $I_{k'\downarrow}$ may be controlled perfectly to flow only by one, two, three or four electron species by varying electric field, exchange field and frequency of circularly polarized light. Very large magnetoresistance dips occur directly related to topological phase transitions. Our work reveal potential of silicene as a topological material for application of spin-valleytronics.

## Acknowledgments

This work is financially supported by Kasetsart University Research and Development Institute (KURDI) and Thailand Research Fund (TRF) under Grant. No. RSA5980058.

| | $C$ | $2C_s$ | $C_v$ | $2C_{ps}$ |
|---|---|---|---|---|
| QAH | $\pm 2$ | 0 | 0 | 0 |
| QSH | 0 | $\pm 2$ | 0 | 0 |
| QVH (trivial insulator) | 0 | 0 | $\pm 2$ | 0 |
| QPSH (trivial insulator) | 0 | 0 | 0 | $\pm 2$ |
| SQAH | $\mp 1$ | $\pm 1$ | $\pm 1$ | $\pm 1$ |
| SQAH | $\pm 1$ | $\mp 1$ | $\pm 1$ | $\pm 1$ |
| SQAH | $\pm 1$ | $\pm 1$ | $\mp 1$ | $\pm 1$ |
| SQAH | $\pm 1$ | $\pm 1$ | $\pm 1$ | $\mp 1$ |

**Table I :** Topological phase transitions in silicene may be described by the first Chern numbers $C$, spin Chern numbers $C_s$, valley-Chern numbers $C_v$ and pseudospin Chern numbers $C_{ps}$. Trivial insulators, QVH and QPSH insulators, are corresponding to zero values of the first and the second Chern numbers $(C, C_s) = (0, 0)$.

**Figure captions**

**Figure 1** Schematic illustration of a silicene-based N/TB/N/TB/N junction where N and TB are normal silicene and topological-phase-transition barriers, respectively (a). Electronic band structure in each region where the Fermi energy in the barriers lies inside band gap to cause topological phase transition (b). In the barriers, proximity-induced exchange energies in A and B sublattices, electric field and irradiating circularly polarized light are applied. Staggered magnetizations in TB-regions can be set to be P-junction and AP-junction.

**Figure 2** Plot of spin-valley conductances as a function of electric field for $E \cong \Delta_{so}$ and $L = 0$. The conductance spectrum, for $\Delta_M = \Delta_\Omega = 0$ (a) plot of $G_{k\uparrow}$ and $G_{k\downarrow}$ by varying thickness of the barriers (b).

**Figure 3** Plot of spin-valley dependent conductances as a function of electric field for $E \cong \Delta_{so}$ and $L = 0$. The conductance spectrum, for P-junction with $\Delta_M \neq 0$, and $\Delta_\Omega = 0$ (a) for $\Delta_M = 0$ and $\Delta_\Omega \neq 0$ (b) and for $\Delta_M = \Delta_\Omega \neq 0$ (c). The junction with current carried by single and two electron groups is predicted.

**Figure 4** Spin-valley-dependent conductances as a function of electric field in AP-junction, for $\Delta_\Omega = 0$ and $\Delta_\Omega \neq 0$ (a) and for $\Delta_M = \Delta_\Omega \neq 0$ (c). Conductance for P and AP-junction in case of $\Delta_M = \Delta_\Omega = \Delta_{so}$ (e). Magnetoresistance as a function of electric field for $\Delta_\Omega = 0$ and $\Delta_M \neq 0$ (b), for $\Delta_M = \Delta_\Omega \neq 0$ (d) and for $\Delta_M = \Delta_\Omega = \Delta_{so}$ (f). All figures are plotted with $L = 0$ and $E \cong \Delta_{so}$. In fig.4(e), currents can be controlled to flow with three-group of electron, $I_{k\downarrow}, I_{k'\uparrow}$ and $I_{k'\downarrow}$, or by single group of electron $I_{k\uparrow}$.

**Figure 5** Plot of conductance as a function of electric field for $\Delta_\Omega = 0$, $\Delta_M \neq 0$, $E \cong \Delta_{so}$ and $2d = 100 \, \text{nm}$ with various values of L, for P-junction (a) and AP-junction (b). Numbers of magnetoresistance dips for $L = 100 \, \text{nm}$ (d) are induced not only due to topological phase transition but also quantum interference.

**Figure 6** Conductance as a function of irradiating circularly polarized light induced-gap $\Delta_\Omega$, for $\Delta_E = \Delta_M = 0$ (a), for $\Delta_E = 0$ and $\Delta_M < \Delta_{so}$ (b), for , for $\Delta_E = 0$ and $\Delta_M > \Delta_{so}$ (c), and for $\Delta_M = \Delta_E = \Delta_{so}$ (d). All figures are plotted with $L = 0$, $2d = 100 \, \text{nm}$ and $E \cong \Delta_{so}$. In Fig.6(d), currents can be controlled to flow with three-

group of electron, $I_{k\uparrow}, I_{k'\uparrow}$ and $I_{k'\downarrow}$, or by single group of electron $I_{k\downarrow}$, different from that in Fig.4(e).

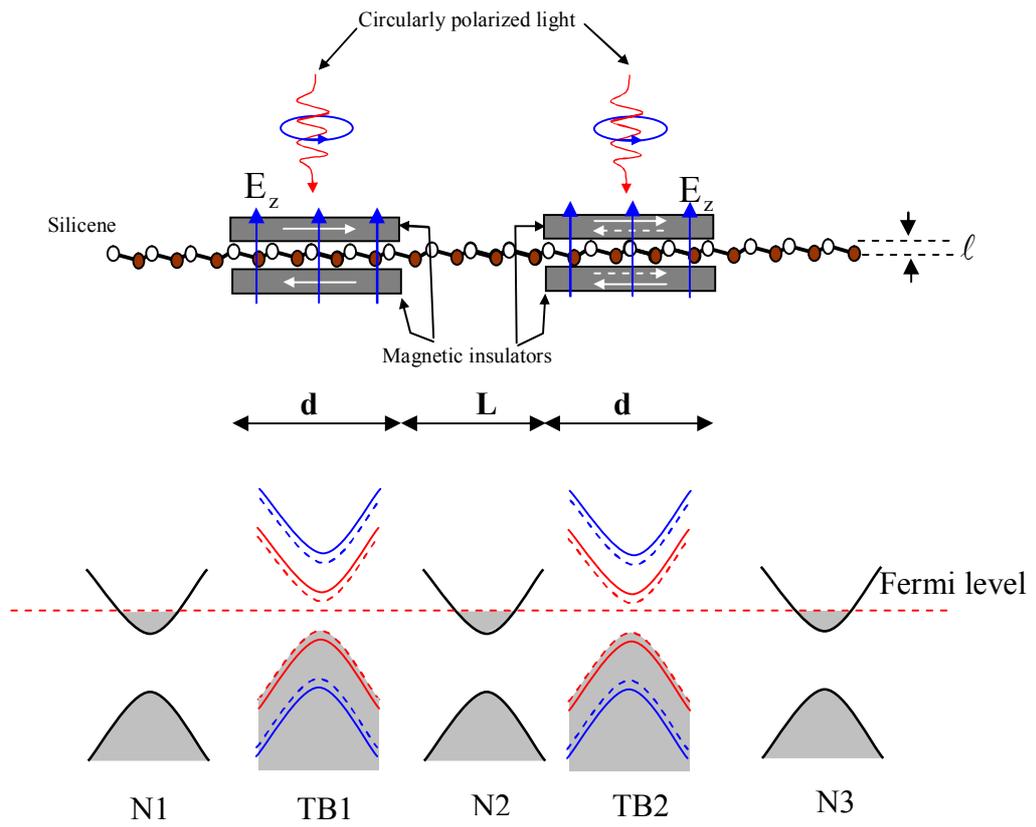

**Figure 1**

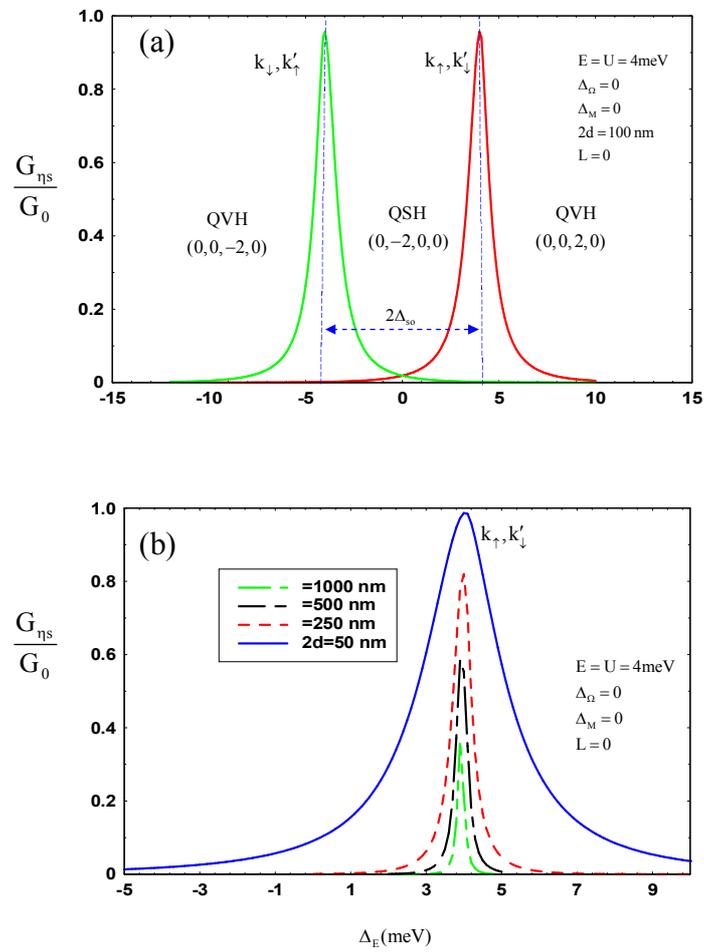

**Figure 2**

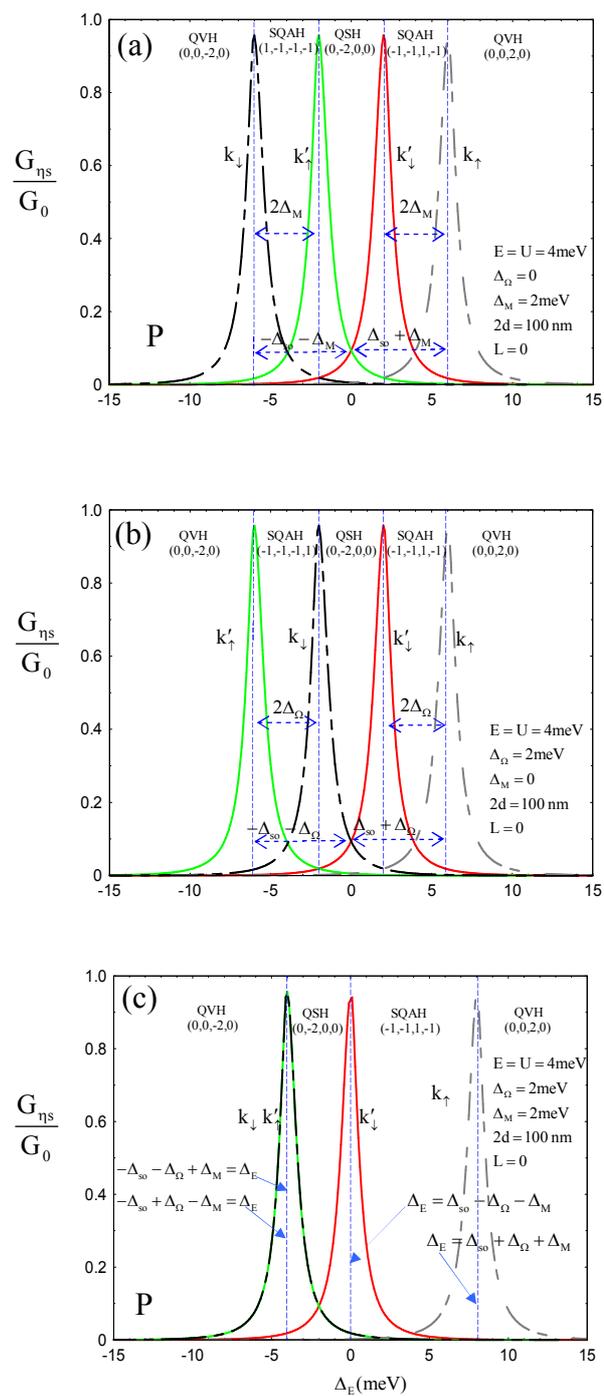

**Figure 3**

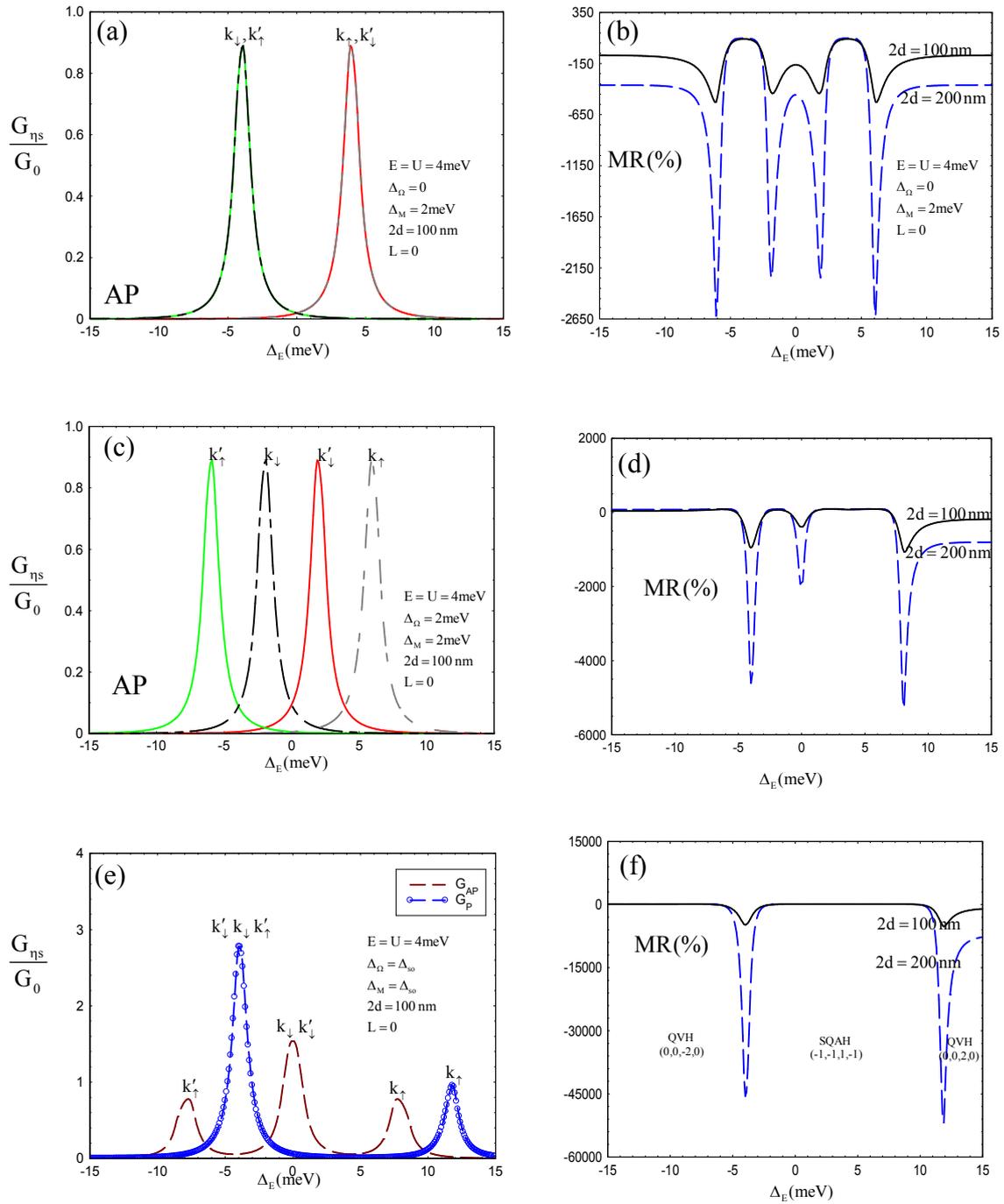

**Figure 4**

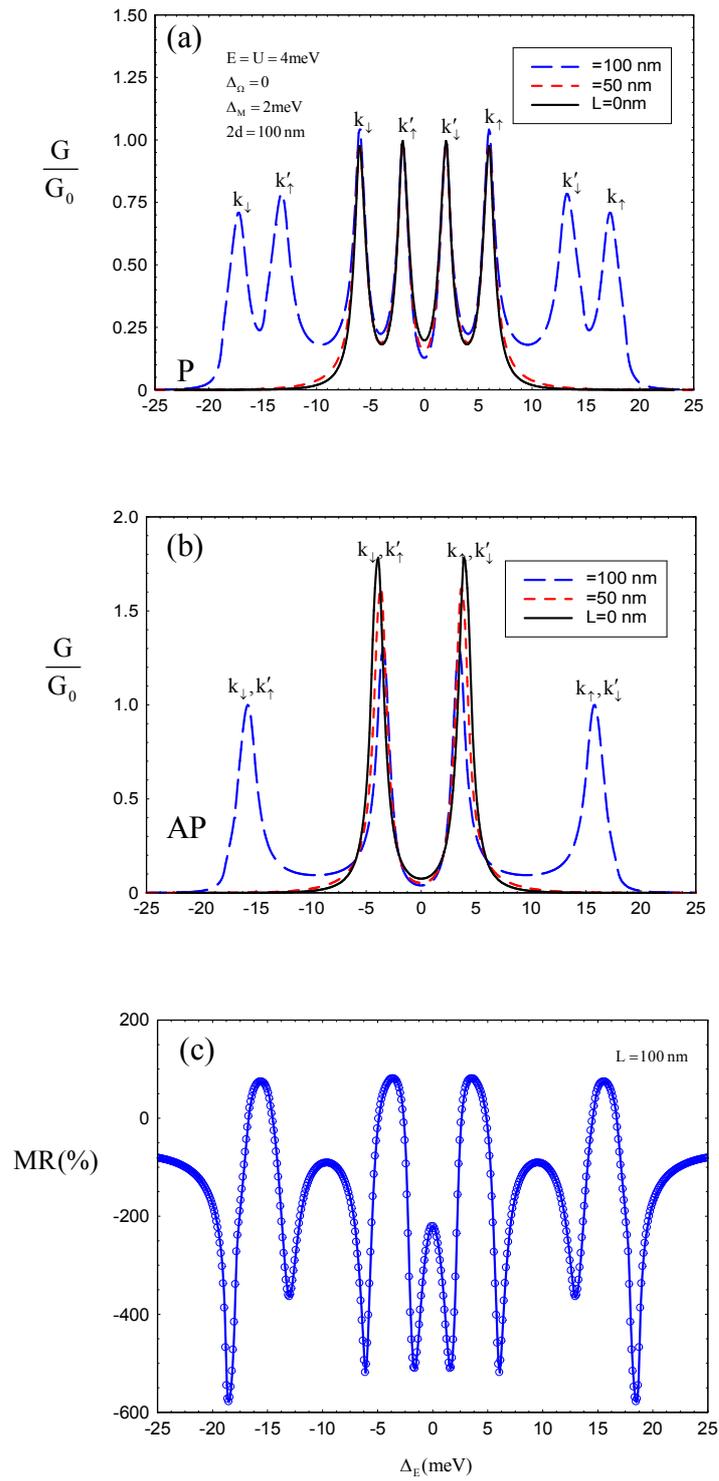

**Figure 5**

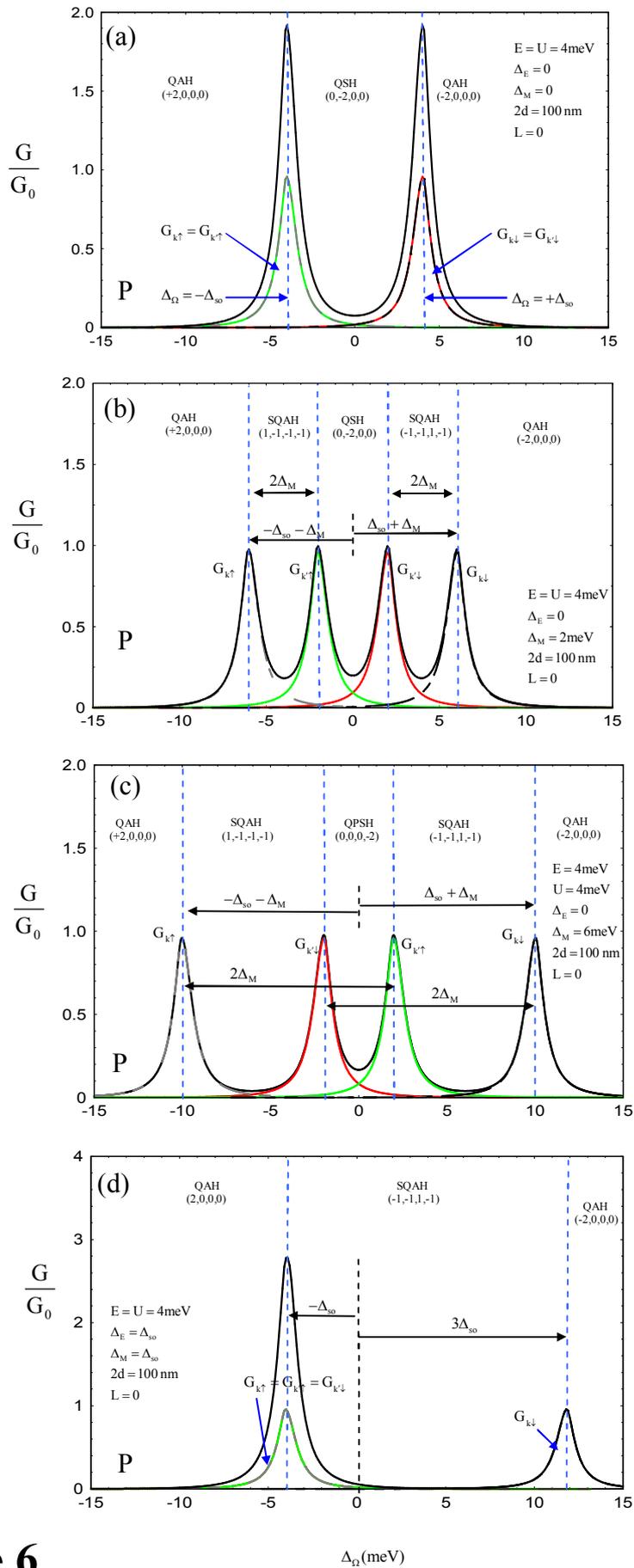

**Figure 6**